\documentclass[10pt,letterpaper]{article}
\usepackage[top=0.85in,left=2.75in,footskip=0.75in]{geometry}

\usepackage{amsmath,amssymb}

\usepackage{changepage}

\usepackage[utf8x]{inputenc}

\usepackage{textcomp,marvosym}

\usepackage{cite}

\usepackage{nameref,hyperref}

\usepackage[right]{lineno}

\usepackage{microtype}
\DisableLigatures[f]{encoding = *, family = * }

\usepackage[table]{xcolor}

\usepackage{array}

\usepackage{arydshln}
\usepackage{multirow}

\newcolumntype{+}{!{\vrule width 2pt}}

\newlength\savedwidth

\raggedright
\usepackage[aboveskip=1pt,labelfont=bf,labelsep=period,justification=raggedright,singlelinecheck=off]{caption}

\makeatletter
\renewcommand{\@biblabel}[1]{\quad#1.}
\makeatother

\date{}

\usepackage{lastpage,fancyhdr,graphicx}
\usepackage{epstopdf}
\fancyheadoffset[L]{2.25in}
\fancyfootoffset[L]{2.25in}
\begin{document}
\vspace*{0.2in}

\begin{flushleft}
{\Large
\textbf\newline{Diversity of Artists in Major U.S. Museums} 
}
\newline
\\
Chad M. Topaz\textsuperscript{1*},
Bernhard Klingenberg\textsuperscript{1,2},
Daniel Turek\textsuperscript{1},
Brianna Heggeseth\textsuperscript{1,3},
Pamela E. Harris\textsuperscript{1},
Julie C. Blackwood\textsuperscript{1},
C.~Ondine Chavoya\textsuperscript{4},
Steven Nelson\textsuperscript{5},
Kevin M. Murphy\textsuperscript{6}
\\
\bigskip
\textbf{1} Department of Mathematics and Statistics, Williams College, Williamstown, MA, USA
\\
\textbf{2} Graduate Program in Data Science, New College of Florida, Sarasota, FL, USA
\\
\textbf{3} Department of Mathematics, Statistics, and Computer Science, Macalester College, St. Paul, MN, USA
\\
\textbf{4} Department of Art, Williams College, Williamstown, MA, USA
\\
\textbf{5} Department of Art History, University of California, Los Angeles, CA, USA
\\
\textbf{6} Williams College Museum of Art, Williamstown, MA, USA
\bigskip

*cmt6@williams.edu

\end{flushleft}
\section*{Abstract}
The U.S. art museum sector is grappling with diversity. While previous work has investigated the demographic diversity of museum staffs and visitors, the diversity of artists in their collections has remained unreported. We conduct the first large-scale study of artist diversity in museums. By scraping the public online catalogs of 18 major U.S. museums, deploying a sample of 10,000 artist records comprising over 9,000 unique artists to crowdsourcing, and analyzing 45,000 responses, we infer artist genders, ethnicities, geographic origins, and birth decades. Our results are threefold. First, we provide estimates of gender and ethnic diversity at each museum, and overall, we find that 85\% of artists are white and 87\% are men. Second, we identify museums that are outliers, having significantly higher or lower representation of certain demographic groups than the rest of the pool. Third, we find that the relationship between museum collection mission and artist diversity is weak, suggesting that a museum wishing to increase diversity might do so without changing its emphases on specific time periods and regions. Our methodology can be used to broadly and efficiently assess diversity in other fields.


\section*{Introduction}
Historically, the artists represented in U.S. art museums have been predominantly male and caucasian. In partnership with the Andrew W. Mellon Foundation, the Association of Art Museum Directors (AAMD) found that 72\% of staff at its member institutions identify as white \cite{SchWes2015}. This same study found that 60\% of museum staff are women, though only 43\% of directorships are held by women, indicating a gender gap at the highest levels of leadership \cite{GanVosPhi2014}. The availability of demographic data has prompted parts of the museum sector to think more intentionally about diversity and inclusion not just amongst staff, but also visitorship. For example, the AAMD has tracked museums' efforts to engage with audiences previously neglected by outreach and education programs, and has helped museums to analyze the geography of visitor origination \cite{AAMD2016}.

All of these diversity efforts involve programs and people rather than collections. If museums find knowledge of staff and visitor demographics important for programming decisions, one might ask if demographics of the artists are important for collection decisions. Anecdotal evidence suggests that some museums are attempting to remediate low levels of diversity in their collections. For instance, in one field, the Art of the United States (U.S.), museums across the country have worked in recent years to incorporate art by women, African Americans, Latinxs, Asian Americans, and Native Americans into a narrative once dominated by white male artists \cite{Hut2003,Wal2010,You2017}. With such increased attention, it is now not unusual for these museums to compete with each other for major works of African American art \cite{Ken2015}.

Efforts towards diversifying collections might be improved by first quantifying the diversity of U.S. art museum collections. We have carried out, and now report on, the first large-scale data science study to measure the genders, ethnicities, geographic origins, and birth decades of artists in the collections of 18 major U.S. art museums.

Our work lies in the context of other studies of diversity in academia, advertising, and employment such as \cite{QuiPagHex2017,NitHebAsh2018,GarMitCue2018}, and has implications for shaping a museum's collection practices and priorities as well as on the disciplines of art history, cultural history  and museum studies. Additionally, beyond the sphere of art, our methodology can be used to assess diversity on a large scale in other fields.

To understand and interpret the diversity of artists in major U.S. art museum collections, it is necessary to study a large data set. While one approach would be to gather artist demographics manually, given the large number of artists, this would be slow and inefficient. Instead, in the spirit of work such as \cite{TopSen2016}, we adopt a crowdsourcing approach and use data science tools to obtain a large dataset of artists and their demographics.

\section*{Methods}

\subsection*{Overview}

We scraped the public online collections of 18 major U.S. art museums, retaining the museum name, artist name, and a web link pointing to the artist's entry in the museum's collection. Then, we deployed a large random subset of scraped records to the crowdsourcing platform Amazon Mechanical Turk (MTurk) and asked crowdworkers to research the demographics of each sampled artist, using the web link as a starting point. Crowdworkers reported their inferences of gender, ethnicity, national origin, and birth year, along with a numerical rating of confidence in each inference. We put in place multiple safeguards and checkpoints to ensure the quality of our data. Starting with the data obtained from crowdworkers, we first eliminated records that did not correspond to individual, identifiable artists. For each remaining record, we aggregated the inferences of multiple crowdworkers and if their responses were sufficiently self-consistent, we made a final inference of each artist's demographic characteristics.

\subsection*{Museum selection}

We started with a convenience sample of 18 museums whose full collections were listed on a publicly accessible website, and that the art and art history scholars on our research team considered to be well-known and representative of the larger U.S. museum landscape. In considering these 18 museums, our art and art history scholars also sought to represent museums that come from different regions of the country and that have different funding models; see columns 2 - 4 of Table~\ref{Table1}. For concision, we henceforth refer to museums by their abbreviations as given in this table. The four geographic regions are those used by the U.S. Census Bureau. The funding models refer specifically to the museum's collection. For example, the MMA building is owned by New York City but the collections are held by a private corporation, so we classified this museum as private.

\begin{table*}[t!]
\begin{adjustwidth}{-2.25in}{0in} 
\caption{Data for the 18 major U.S. art museums in our study. These museums are distributed across U.S. geographic regions and have a range of collection funding model types. Museums are arranged into groups according to their collecting mission as determined by a cluster analysis; see Results. The table below summarizes our data, including the date we scraped records from websites, the number of records scraped, the number of records randomly sampled from the scraped records, and the number and percentage of sampled records from a museum's collection determined to be individual, identifiable artists (IIA). Overall, there are 10,108 IIA records and we make confident gender, ethnicity, regional origin, and birth decade inferences (CGI, CEI, CRI, CBI) for, respectively, 89\%, 82\%, 83\%, and 79\% of these.\label{Table1}}
\small
\setlength\tabcolsep{2pt}
\begin{tabular*}{\linewidth}{@{\extracolsep{\fill}}clcccrrrrrrr}
\hline
Group & Museum$^1$ & Region$^2$ & Type$^3$ & Date & Scraped & Sampled & \multicolumn{1}{c}{IIA} &  \multicolumn{1}{c}{CGI} & \multicolumn{1}{c}{CEI} & \multicolumn{1}{c}{CRI} & \multicolumn{1}{c}{CBI}\\ \hline
{\multirow{5}{*}{1}}&DIA&M&PUB&2017-06-28&5528&751&627 (83\%)&578 (92\%)&509 (81\%)&523 (83\%)&506 (81\%)\\
&MMA&N&PRI&2018-05-11&35612&764&669 (88\%)&589 (88\%)&531 (79\%)&517 (77\%)&411 (61\%)\\
&MFAB&N&PRI&2017-07-17&24994&785&611 (78\%)&503 (82\%)&472 (77\%)&471 (77\%)&406 (66\%)\\
&NGA&S&PUB&2017-06-28&13856&400&374 (94\%)&336 (90\%)&309 (83\%)&325 (87\%)&295 (79\%)\\
&PMA&N&PRI&2017-06-28&14329&747&654 (88\%)&560 (86\%)&533 (81\%)&520 (80\%)&462 (71\%)\\\hdashline
{\multirow{4}{*}{2}}&AIC&M&PRI&2017-06-28&10998&466&405 (87\%)&359 (89\%)&342 (84\%)&352 (87\%)&330 (81\%)\\
&NAMA&M&PRI&2017-06-28&4725&627&570 (91\%)&509 (89\%)&472 (83\%)&463 (81\%)&434 (76\%)\\
&RISDM&N&UNI&2018-05-08&4283&780&620 (79\%)&535 (86\%)&478 (77\%)&498 (80\%)&440 (71\%)\\
&YUAG&N&UNI&2018-05-17&10200&710&668 (94\%)&586 (88\%)&563 (84\%)&569 (85\%)&532 (80\%)\\\hdashline
{\multirow{5}{*}{3}}&DMA&S&PRI&2017-06-29&4053&700&605 (86\%)&551 (91\%)&495 (82\%)&503 (83\%)&482 (80\%)\\
&DAM&W&PRI&2017-06-28&912&776&733 (94\%)&675 (92\%)&610 (83\%)&644 (88\%)&648 (88\%)\\
&HMA&S&PRI&2018-05-10&440&430&402 (93\%)&382 (95\%)&348 (87\%)&365 (91\%)&371 (92\%)\\
&LACMA&W&PUB&2018-05-17&14164&732&635 (87\%)&548 (86\%)&513 (81\%)&552 (87\%)&510 (80\%)\\
&MFAH&S&PRI&2018-05-14&13250&867&696 (80\%)&615 (88\%)&560 (80\%)&570 (82\%)&560 (80\%)\\\hdashline
{\multirow{3}{*}{4}}&MOCA&W&PRI&2017-06-26&1226&424&419 (99\%)&389 (93\%)&377 (90\%)&376 (90\%)&394 (94\%)\\
&MOMA&N&PRI&2017-06-27&21187&445&376 (84\%)&337 (90\%)&300 (80\%)&286 (76\%)&295 (78\%)\\
&SFMOMA&W&PRI&2017-06-28&3376&596&531 (89\%)&493 (93\%)&450 (85\%)&473 (89\%)&478 (90\%)\\\hdashline
{\multirow{1}{*}{5}}&WMAA&N&PRI&2017-06-02&3524&522&513 (98\%)&471 (92\%)&434 (85\%)&425 (83\%)&472 (92\%)\\\hdashline
& \textbf{Overall}&&&&\textbf{186657}&\textbf{11522}&\textbf{10108 (88\%)}&\textbf{9016 (89\%)}&\textbf{8296 (82\%)}&\textbf{8432 (83\%)}&\textbf{8026 (79\%)}\\
\hline
\end{tabular*}
\\
\footnotesize{$^1$DIA = Detroit Institute of Arts; MMA = Metropolitan Museum of Art; MFAB = Museum of Fine Arts, Boston; NGA = National Gallery of Art; PMA = Philadelphia Museum of Art; AIC = Art Institute of Chicago; NAMA = Nelson-Atkins Museum of Art; RISDM = Museum of Art, Rhode Island School of Design; YUAG = Yale University Art Gallery; DMA = Dallas Museum of Art; DAM = Denver Art Museum; HMA = High Museum of Art; LACMA = Los Angeles County Museum of Art; MFAH = Museum of Fine Arts, Houston; MOCA = Museum of Contemporary Art, Los Angeles; MOMA = Museum of Modern Art; SFMOMA = San Francisco Museum of Modern Art; WMAA = Whitney Museum of American Art.}
\\
\footnotesize{$^2$M = Midwest, N = Northeast, S = South, W = West.}
\\
\footnotesize{$^3$PUB = Public, PRI = Private, UNI = University.}
\normalsize
\end{adjustwidth}
\end{table*}

\subsection*{Data Scraping}

For each museum in our study, we read the museum website Terms of Service and verified that gathering data for research would comply with them, as well as with United States Fair Use doctrine. We then scraped the catalog of holdings using customized code we wrote with the \texttt{R} project for statistical computing \cite{R2017}, retaining a web link for each item \cite{museumsites}. Depending on the museum, the link pointed to a web page of biographical information about the artist, a specific work by the artist, a page depicting multiple works by the artist, or, simply, the museum's main web page.

Every museum catalog in our study contained records that did not correspond to individual, identifiable artists (IIA). Some of these records were for unknown artists appearing with blank information or the phrase ``unknown.'' Some corresponded to works with only geographic, cultural, and/or temporal information available, \emph{e.g.}, ``North American Indian'' or ``Unidentified artist, French, second quarter 18th century.'' Some were for works created by a design, production, or architecture firm, \emph{e.g.}, ``Amstel Porcelain Company.'' Finally, some were works created by a collaboration of individuals, \emph{e.g.}, ``Maurice Loewy and Pierre Puiseux.''

Because much of the scraped data did not carry structured metadata, and because there exist many different types of non-IIA records (as described above), it is difficult to automatically classify artist records as IIA vs.\ non-IIA. At this stage, we only excluded records for which the artist field included ``Company,'' ``\& Co,'' or ``\& Sons.'' For all remaining records, we deferred IIA determination to crowdsourcing.

\subsection*{Survey Instrument}

\subsubsection*{Survey Overview}

Some museums in our study listed online artist birth years and geographic origins. Of the 18 museums we selected, only one, namely NAMA, provided any gender information, and this information was only available for a subset of the artists. No museum websites listed artist ethnicity. We inferred all four demographic traits through a survey instrument deployed via crowdsourcing. The instrument was drafted, tested, and refined in a small pilot study of 600 artist-museum records randomly sampled from four museums in our study. In our final instrument, crowdworkers were provided with an artist-museum record. Workers were instructed to research the artist using the museum web link provided and any other resources, potentially including Google and Wikipedia. The instrument consisted of 10 questions; the complete text of our survey appears in \cite{HegKli2018}. The first question asked whether the artist is an IIA. All subsequent questions allowed the worker to answer that the artist is non-IIA. These subsequent questions asked for the artist's gender (including an option for nonbinary), primary ethnicity (allowing for more than one), national origin, and birth year. For each demographic question, we also asked workers for their degree of confidence in their response (low~=~1/3, medium~=~2/3, high~=~3/3).

\subsection*{Sampling and Crowdsourcing}

We deployed our survey instrument via MTurk. MTurk is a virtual labor market for Human Intelligence Tasks (HITs), tasks that are difficult or impossible to automate with computer code. In the MTurk system, we acted as requesters, posting HITs that workers could complete for wages that we set. MTurk is commonly used for research in the social sciences and computer science \cite{PhaAir2015,GalYan2015}. Because of its growing importance as a research tool, MTurk has itself become a subject of news and study \cite{Dan2015}.

When deploying HITs to MTurk, we took six steps to ensure data quality. First, we only hired workers who had completed at least 1,000 HITs prior to participating in our study, guaranteeing a sufficient level of experience. Second, we only hired workers with an approval rating of 99\% or higher for past work, providing confidence in their responses. Third, and most importantly, we had each artist-museum record researched by at least five independent workers. As we will describe, we used this redundancy to form consensus demographic inferences. Fourth, fifth and sixth, as we will also describe, we removed data from workers who appeared not to participate in good faith, manually validated a subsample of our demographic inferences, and checked internal and external consistency of the data.

We deployed HITs in two stages. In the first stage, we randomly sampled 400 records from each museum's scraped artist database. The second stage sample size varied from museum to museum based on the proportion of IIAs determined from the first stage. Those museums with a high proportion of IIAs needed fewer additional samples to achieve our target margin of error of 3.3\% for gender inference compared to museums with smaller proportions of IIAs. The 3.3\% margin of error was based on demographic information from our pilot study and budgetary restrictions. The combined stage~1 and stage~2 sample size of artist records drawn from each museum is shown in Column 7 of Table~\ref{Table1}. Due to investigator error, 279 artist-museum records were deployed more than once; we retained the unintentionally gathered data. In total, we deployed 59,630 HITs to MTurk, corresponding to 11,522 unique artist-museum pairs.

A small number of workers completed unusually many HITs, with most completed in 30 seconds or less. We analyzed data submitted by the 17 most prolific workers, who each completed between 581 and 1,391 HITs, and found that eight of them appear to have acted in bad faith, submitting large quantities of incorrect information. We excluded all data from these eight workers, totaling 7,147 HITs, or 12\% of the HITs deployed. This does not significantly decrease our ability to analyze artist demographics, since we deployed at least five HITs per record. After removing data from bad-faith workers, only 107 of our artist-museum records (less than 1\%) had fewer than three HITs. Nonetheless, for future studies, we would limit the number of HITs a worker can complete in order to protect against dishonest behavior. For insights into the reliability of raw data gathered on MTurk, see, \emph{e.g.}, \cite{PeeVosAcq2014,Rou2015}. After eliminating data from the bad-faith workers, we calculated an average HIT completion time of 106 seconds, resulting in an average hourly wage of just over \$10 paid to workers.

\subsection*{Individual, Identifiable Artists}

We used responses to Question 1 of our survey to evaluate which records corresponded to IIA. To identify a record as IIA, we required that it was evaluated by at least three workers, and that a majority of the workers classified it as IIA. We excluded HIT results for which the worker disagreed with the IIA consensus. For example, for \'{E}lizabeth Baillon from PMA's collection, four out of five workers indicated that the artist was IIA. We therefore retained the record as IIA but excluded the one worker who disagreed. Column 8 of Table~\ref{Table1} shows the number and percentage of IIA records we identified for each museum. Overall, we found 10,108 records to be IIA, representing 88\% of those sampled and comprising 45,202 HITs. We restricted attention to these data, which we refer to as IIA records and IIA responses. For each demographic inference below, we excluded IIA responses for which the worker indicated that they could not determine an inference or that the artist was not IIA (inconsistent with their Question 1 response), resulting in slightly fewer than 45,202 completed HITs. For each demographic inference below, we state the number of available responses.

\subsection*{Gender}

Of 43,576 IIA responses available for gender analysis, 54 identified an artist's gender as nonbinary, each corresponding to a different IIA record. We excluded these responses from gender analysis as we could not confidently conclude any artists to have nonbinary gender. For each remaining response, we created a gender score by assigning the value -1 to man and +1 to woman, and weighting by the degree of confidence \cite{TopSen2016}. We then averaged all gender scores for a given record. To make a confident gender inference, we required the mean gender score of an IIA record to be at least $0.65$ in magnitude (see Validation) and we required at least three IIA responses. We refer to cases meeting these requirements as confident gender inferences (CGI). For example, artist Bai Yiluo had four responses inferring gender. Three inferred man with high confidence 3/3 and one inferred woman with low confidence 1/3, so the mean gender score was $(1/4)[(-1)\cdot 3/3 + (-1)\cdot 3/3 +(-1)\cdot 3/3 + (1)\cdot 1/3]= -0.67$, and we inferred the artist to be a man. Column~9 of Table~\ref{Table1} summarizes CGI records.

\subsection*{Ethnicity}

Of 41,353 IIA responses available for ethnicity analysis, only 592 provided a free-text ethnicity. Nearly all of the free-text responses erroneously contained birth year or national origin information. We excluded all free-text data but retained multiple choice responses since workers could also select ethnicities from our list. This list consisted of race and ethnicity categories from the U.S. census augmented with a category for Middle Eastern or North African people, who would otherwise be classified as white under the strict census scheme. Most ethnicities in our data set were Asian, Black/African American, Hispanic/Latinx or white. Only 2,903 responses indicated the remaining ethnicities in our list, namely American Indian or Alaska Native, Native Hawaiian or other Pacific Islander, and Middle Eastern or North African. We combined these three categories into one called ``other.'' For each IIA record and each ethnicity, we assigned a mean confidence-weighted score similar to our gender score, but now ranging between 0 and 1, with~1 corresponding to a strong consensus that the artist is a member of the ethnic group. If the mean score exceeded 0.65 and there were at least three IIA responses indicating that ethnicity, then we inferred that the artist had that ethnicity. Though our survey instrument allowed for the possibility of multiple primary ethnicities, we found that only two out of our 10,108 IIA records were confidently inferred as such. We excluded these two IIA records and treated primary ethnicity as a single categorical variable with five categories. Our confident ethnicity inferences (CEI) are summarized in Column 10 of Table~\ref{Table1}.

\subsection*{Geographic Origin}
For the 42,253 IIA responses available for geographic analysis, we mapped national origin response to one of the seven GEO3 regions used by the United Nations \cite{unep2002}. For each IIA record, we assigned a mean confidence-weighted score to each region attributed to the artist. If the mean score was greater than or equal to 0.8 and there were at least three IIA HITs for that region, we accepted that region as a confident regional origin inference (CRI); otherwise we made no inference. For example, for artist Adriane Herman, four workers chose national origin of USA with confidence 3/3 and one chose Denmark with confidence 3/3. We assigned a score of $(1/5)[3/3+3/3+3/3+3/3]=0.8$ to North America and $(1/5)(3/3)=0.2$ to Europe, and inferred the region as North America. Column 11 of Table~\ref{Table1} summarizes CRI records.

\subsection*{Birth Decade}
Of the IIA HITs available for analysis of birth year, we retained the 40,424 responses containing three or four digit numbers. Within each artist-museum record, we calculated a $z$-score for each response and discarded responses with $|z|>1$, unless the response was within one year of the mean, in which case we accepted it as a small typographical error. If an IIA record had at least three responses after discarding outliers, we made a confident birth decade inference (CBI) by taking a confidence-weighted average of the birth years and rounding to the nearest decade. For example, for artist Alonso S\'{a}nchez Coello, one worker replied with an empty birth year, two chose 1831 with highest confidence, and two chose 1832 with highest confidence. We discarded the empty response and computed the average birth year $(1/4)[(3/3) \cdot 1831 + (3/3) \cdot1831 + (3/3) \cdot 1832 + (3/3) \cdot 1832]$, which to the nearest decade is 1830. Column 12 of Table~\ref{Table1} summarizes CBI records.

\subsection*{Validation}
Above, we set score thresholds to make confident inferences, for example, 0.65 on our gender score. We chose these thresholds to give agreement with data we validated. To validate crowdsourced inferences, our team researched demographics for 374 randomly sampled artist-museum records, with at least 20 sampled per museum. This validation set accounted for 3.7\% of our IIA records. We were able to manually infer gender for 368 records, geographic origin for 370, and birth decade for 350, and obtained perfect agreement, with no errors in the crowdsourced inferences. For ethnicity, we made 291 inferences and found one discrepancy. We inferred Rirkrit Tiravanija (in the MCA collection) to be Asian because his parents are Thai, whereas crowdworkers chose Hispanic/Latinx, likely because he was born in Bueno Aires. This error accounts for 0.3\% of the validated data on ethnicity.

Because several artists (717 in total) appeared in two or more collections, we could check the internal consistency of our data to see if an artist's demographics were consistently inferred across collections. For gender, we found no inconsistencies, although for 79 artists we found a confident gender inference in some museums but not others. For instance, Rachel Lachowicz was inferred as a woman in WMAA and MOCA but had missing gender inference in DAM. For such cases, we verified gender and filled in the missing information. For ethnicity, we had four inconsistencies. For example, Paul Pfeiffer was correctly inferred as Other in SFMOMA but as white in RISDM. We manually corrected these inconsistencies, and similarly, we corrected two inconsistencies for geographic origin and five for birth decade.

To check external consistency, we compared our crowdsourced gender inferences to gender information posted on NAMA's website. NAMA is the only one of the 18 museums that provides gender information on some of their artists. Our sample included 570 IIA records from NAMA. Out of these, NAMA's website had 292 artists (51\%) with no gender specified, 198 artists (35\%) specified as men, and 80 artists (14\%) specified as women. Of the 292 artists with no gender label, our crowdsourcing process confidently inferred 257 (88\%) to be men and 5 (2\%) to be women. For the remaining 30 artists (10\%), crowdsourcing was also not able to provide confident gender inference. Of the 198 artists that NAMA identified as men, crowdsourcing inferred 191 (96\%) as men, zero as women, and 7 (4\%) as unknown (no confident inference possible). Of the 80 artists that NAMA identified as women, crowdsourcing inferred one (1\%) as a man, 55 (69\%) as women, and could not confidently infer gender for the remaining 24 (30\%) of the artists. The one artist (Maxime Du Camp) NAMA labeled as a woman and crowdsourcing inferred as a man is in fact a man, based on the artist's Wikipedia entry.

Finally, we checked our dataset against another independent but also crowdsourced (and hence possibly erroneous) dataset. The website \texttt{www.the-athenaeum.org/} publishes a database of roughly 1,000 women artists with records contributed by users of that site. This database contains 68 of the artists that we also sampled for our study. Of these, our crowdsourcing process inferred one (1\%) as a man (Vivian Forbes) and, based on the artist's Wikipedia entry, it appears that this inference is correct. Our crowdsourcing inferred 43 artists (63\%) as women and for the remaining 11 (16\%), it could not confidently infer gender.

These internal and external validations provide confidence into our inferences, but they also show that recognizing women artists as such may be more difficult than recognizing men using our crowdsourcing approach. This difficulty mirrors the broader issues of systemic bias and women's reduced prominence on the Internet; see, \emph{e.g.}, \cite{LamUduDon2011}. As a consequence, the true proportion of women artists in our study may be slightly larger than what we report here. Below, we report demographic proportions explicitly accounting for statistical sampling error, but we are aware that a small unknown margin of crowdsourcing error may also be present.

\section*{Results}

The artist dataset described above can be accessed and explored online via a web app \cite{HegKli2018}. Our analysis can be reproduced via R code in a GitHub repository \cite{github}.

Having inferred and validated demographic data, we now distinguish between two types of demographics. We interpret gender and ethnicity as demographics reflective of artist diversity, and we interpret regional origin and birth decade as reflective of a museum's collection mission and priorities. Below, we present three types of results: (1) overall levels of artist diversity at each museum, (2) identification of outliers that have significantly more or less demographic diversity than the pool of museums at large, and (3) levels of artist diversity interpreted vis-a-vis museum collection mission demographics. In referring to specific museums, for concision, we again use the abbreviations of museum names as given in Table~\ref{Table1}.

\subsection*{Diversity Measurements} 

Using the confident inferences made by crowdsourcing, we estimated the gender and ethnicity distributions overall and for each museum, summarized in Table~\ref{Table2}. In the following, we express all demographic proportions with respect to the pool of individual, identifiable artists for whom we could reach confident demographic inference. In computing statistics for the overall pool (last row of Table~\ref{Table2}) we count artists that appear in multiple museums only once. To identify such artists, we used a perfect match in the artist name, leaving us with 9,188 unique artist names. Some artists still may be represented more than once if their name is spelled differently in different museums. However, the effect of this on the overall proportions should be negligible. 

With respect to gender, our overall pool of individual, identifiable artists across all museums consists of 12.6\% women. With respect to ethnicity, the pool is 85.4\% white, 9.0\% Asian, 2.8\% Hispanic/Latinx, 1.2\% Black/African American, and 1.5\% other ethnicities. The four largest groups represented across all 18 museums in terms of gender and ethnicity are white men (75.7\%), white women (10.8\%), Asian men (7.5\%), and Hispanic/Latinx men (2.6\%). All other groups are represented in proportions less than 1\%. To give a sense of the sampling error, Table~\ref{Table2} shows (simultaneous) 95\% confidence intervals for the true proportions, computed using the score approach with a Bonferroni adjustment \cite{Agr2003}.

\begin{table*}[t!]
\begin{adjustwidth}{-2.25in}{0in} 
\caption{Demographic diversity results obtained from our data, with museums listed according to their collection mission cluster; see Table~\ref{Table1} and Results. For each museum, we give the  observed percentage of the demographic (gender, ethnicity) as well as a  confidence interval for the true proportion. The colored cells indicate  that a museum's proportion differs significantly from the overall proportion (excluding that museum) at the familywise 5\% significance level. For each demographic, the 18 confidence intervals are multiplicity adjusted via Bonferroni so that the simultaneous coverage probability is 95\%. Similarly, the underlying p-values for the tests comparing proportions are also multiplicity adjusted via Bonferroni to control the familywise error rate at 5\%. \label{Table2}}
\small
\setlength\tabcolsep{6pt}
\begin{tabular*}{\hsize}{@{\extracolsep{\fill}}clrcrcrcrcrcrc}
\cline{1-14}
Group&Museum & \multicolumn{2}{c}{Women} & \multicolumn{2}{c}{Asian} & \multicolumn{2}{c}{Black/Af. Am.} & \multicolumn{2}{c}{Hisp./Lat.} & \multicolumn{2}{c}{White} & \multicolumn{2}{c}{Other}\\
&& \% & 95\% CI  & \% & 95\% CI & \% & 95\% CI & \% & 95\% CI & \% & 95\% CI & \% & 95\% CI \\ \cline{1-14}
{\multirow{5}{*}{1}}&DIA&7.4\cellcolor{red!25}&[4.8,11.4]&2.8\cellcolor{red!25}&[1.1,6.5]&1.6&[0.5,4.9]&0.4&[0.1,3.0]&94.7\cellcolor{red!25}&[90.1,97.2]&0.6&[0.1,3.3]\\
&MMA&7.3\cellcolor{red!25}&[4.7,11.2]&8.1&[4.9,13.2]&0.2&[0.0,2.6]&1.5&[0.5,4.7]&88.9&[83.3,92.8]&1.3&[0.4,4.4]\\
&MFAB&8.2\cellcolor{red!25}&[5.2,12.6]&16.1\cellcolor{green!25}&[11.1,22.8]&1.1&[0.3,4.3]&2.1&[0.7,5.8]&79.9\cellcolor{green!25}&[72.8,85.5]&0.8&[0.2,3.9]\\
&NGA&10.4&[6.4,16.5]&1.3\cellcolor{red!25}&[0.3,5.9]&0.0&[0.0,3.7]&0.6&[0.1,4.9]&97.4\cellcolor{red!25}&[92.1,99.2]&0.6&[0.1,4.9]\\
&PMA&8.8&[5.8,13.0]&8.3&[5.0,13.3]&1.1&[0.3,4.1]&2.4&[1.0,6.0]&87.8&[82.1,91.9]&0.4&[0.0,2.9]\\\hdashline
{\multirow{4}{*}{2}}&AIC&12.5&[8.2,18.7]&7.0&[3.6,13.4]&0.3&[0.0,3.9]&2.0&[0.6,6.7]&90.4&[83.4,94.6]&0.3&[0.0,3.9]\\
&NAMA&11.6&[8.0,16.5]&9.5&[5.8,15.2]&0.4&[0.1,3.2]&1.3&[0.3,4.6]&86.4&[80.1,91.0]&2.3&[0.9,6.1]\\
&RISDM&13.1&[9.3,18.1]&15.1\cellcolor{green!25}&[10.3,21.6]&1.0&[0.3,4.2]&3.1&[1.3,7.2]&78.2\cellcolor{green!25}&[71.1,84.0]&2.5&[1.0,6.4]\\
&YUAG&11.6&[8.2,16.2]&14.2\cellcolor{green!25}&[9.9,20.0]&0.7&[0.1,3.3]&2.3&[0.9,5.7]&81.7&[75.4,86.7]&1.1&[0.3,3.9]\\\hdashline
{\multirow{5}{*}{3}}&DMA&15.1&[11.1,20.2]&4.2\cellcolor{red!25}&[2.0,8.6]&0.8&[0.2,3.8]&2.8&[1.2,6.7]&88.7&[82.8,92.7]&3.4\cellcolor{green!25}&[1.5,7.5]\\
&DAM&13.3&[9.9,17.7]&9.5&[6.1,14.4]&1.5&[0.5,4.3]&5.4\cellcolor{green!25}&[3.0,9.5]&79.8\cellcolor{green!25}&[73.7,84.8]&3.8\cellcolor{green!25}&[1.9,7.4]\\
&HMA&10.7&[6.9,16.4]&0.9\cellcolor{red!25}&[0.1,4.8]&10.6\cellcolor{green!25}&[6.2,17.7]&1.4&[0.3,5.7]&86.2&[78.6,91.4]&0.9&[0.1,4.8]\\
&LACMA&10.6&[7.3,15.2]&17.7\cellcolor{green!25}&[12.7,24.3]&0.0&[0.0,2.3]&2.9&[1.2,6.7]&78.2\cellcolor{green!25}&[71.3,83.8]&1.2&[0.3,4.2]\\
&MFAH&16.1&[12.2,21.0]&4.3\cellcolor{red!25}&[2.2,8.3]&1.1&[0.3,3.9]&4.8&[2.5,9.0]&88.6&[83.1,92.4]&1.2&[0.4,4.2]\\\hdashline
{\multirow{3}{*}{4}}&MOCA&24.9\cellcolor{green!25}&[19.0,32.0]&6.9&[3.6,12.8]&2.7&[0.9,7.3]&6.4\cellcolor{green!25}&[3.2,12.2]&82.8&[75.1,88.4]&1.3&[0.3,5.3]\\
&MOMA&11.0&[6.9,17.1]&10.0&[5.5,17.6]&2.0&[0.5,7.1]&3.7&[1.4,9.5]&83.0&[74.3,89.2]&1.3&[0.3,6.1]\\
&SFMOMA&18.1\cellcolor{green!25}&[13.5,23.8]&7.1&[3.9,12.5]&2.0&[0.7,5.8]&3.3&[1.4,7.7]&86.4&[79.9,91.1]&1.1&[0.3,4.5]\\\hdashline
{\multirow{1}{*}{5}}&WMAA&22.1\cellcolor{green!25}&[16.9,28.3]&2.8\cellcolor{red!25}&[1.1,7.0]&2.3&[0.8,6.3]&2.3&[0.8,6.3]&91.7\cellcolor{red!25}&[85.9,95.2]&0.9&[0.2,4.3]\\\hdashline
&\textbf{Overall}&\textbf{12.6}&\textbf{[11.9,13.4]}&\textbf{9.0}&\textbf{[8.2,9.9]}&\textbf{1.2}&\textbf{[0.9,1.6]}&\textbf{2.8}&\textbf{[2.4,3.4]}&\textbf{85.4}&\textbf{[84.3,86.4]}&\textbf{1.5}&\textbf{[1.2,1.9]}\\
\cline{1-14}
\end{tabular*}
\normalsize
\end{adjustwidth}
\end{table*}

To provide some perspective on these results, we make several comparisons. First, we consider the 2016 American Community Survey \cite{ACS}, an ongoing survey that is run by the U.S. Census Bureau but is distinct from the decennial census. Because artists in our database come from many time periods and geographic regions, a direct comparison with ACS data would not be meaningful. To move towards a more meaningful comparison, we restrict our attention to artist records in our database with geographic origin in North America and with birth year of 1945 or later. In this subset, there were 831 CGIs, with women constituting 29.8\%. In contrast, the ACS shows the current gender breakdown of artists and related workers to be 43.3\% women. From our artist database, we could make 724 CEIs, with 91.7\% white, 3.9\% Black/African American, 1.0\% Hispanic/Latinx, 0.6\% Asian, and 2.9\% other ethnicities. The ACS found the most common single races/ethnicities of artist to be white (78.6\%), Asian (11.5\%), and Black/African American (2.3\%). We can also compare our results to the general U.S. population via the decennial census, which found the U.S. population to be 50.9\% women, 72.4\% white, 12.6\% Black/African American, and 4.8\% Asian \cite{census1,census2}. Hence our artist database is more white and male as compared to ACS and census data. It is important to note that the ACS uses census demographic information, which separates the question of race and ethnicity. This means that individuals must select a race independent of whether they identify ethnically as Hispanic/Latinx. If one accounts for the change in distributions using white non-Hispanic/Latinx, then the U.S. population is 69.1\% white \cite{census3}.

\subsection*{Diversity Outlier Analysis}

To detect museums with significantly different diversity from our pool of data at large, we performed statistical tests for differences in proportions. Specifically, we compared a museum's demographic proportion to the remainder of the pool, excluding that museum. Colored cells in Table~\ref{Table2} indicate instances in which the museum's proportion was significantly different from the pool using a multiplicity adjusted p-value less than $0.05$. Red cells indicate less diversity (more male, less nonwhite, or more white) and green cells indicate greater diversity.

With respect to gender, DIA, MMA, and MFAB have a significantly lower proportion of women, while MOCA, SFMOMA, and WMAA are higher. With respect to the proportion of Asian artists, DIA, NGA, DMA, HMA, MFAH, and WMAA are low and  MFAB, RISDM, YUAG, and LACMA are high. For Black/African American artists, no museums are significantly lower than the overall level, while HMA stands out with high representation. Similarly, for Hispanic/Latinx artists, we observe outliers only on the high end, namely DAM and MOCA. Museums with higher representation of white artists are DIA, NGA, and WMAA, whereas MFAB, RISDM, DAM, and LACMA are less white. Finally, in the ``other'' ethnicity category, there are no outliers on the low end, and DMA and DAM have higher proportions than the pool.

\subsection*{Diversity and Museum Collection Mission}

Museums have different collection missions characterized, typically, by geographic regions and by time periods. Most of the museums we selected are considered encyclopedic in their collecting missions, aspiring to represent the whole of human visual expression across geography and time. Table~\ref{Table3} shows the geographic origin proportions we inferred for each museum as well as birth decade information summarized by the average over all IIA records within each museum after removing duplicates.

\begin{figure}
\begin{adjustwidth}{-2.25in}{0in}
\centering
\includegraphics[width=1\textwidth]{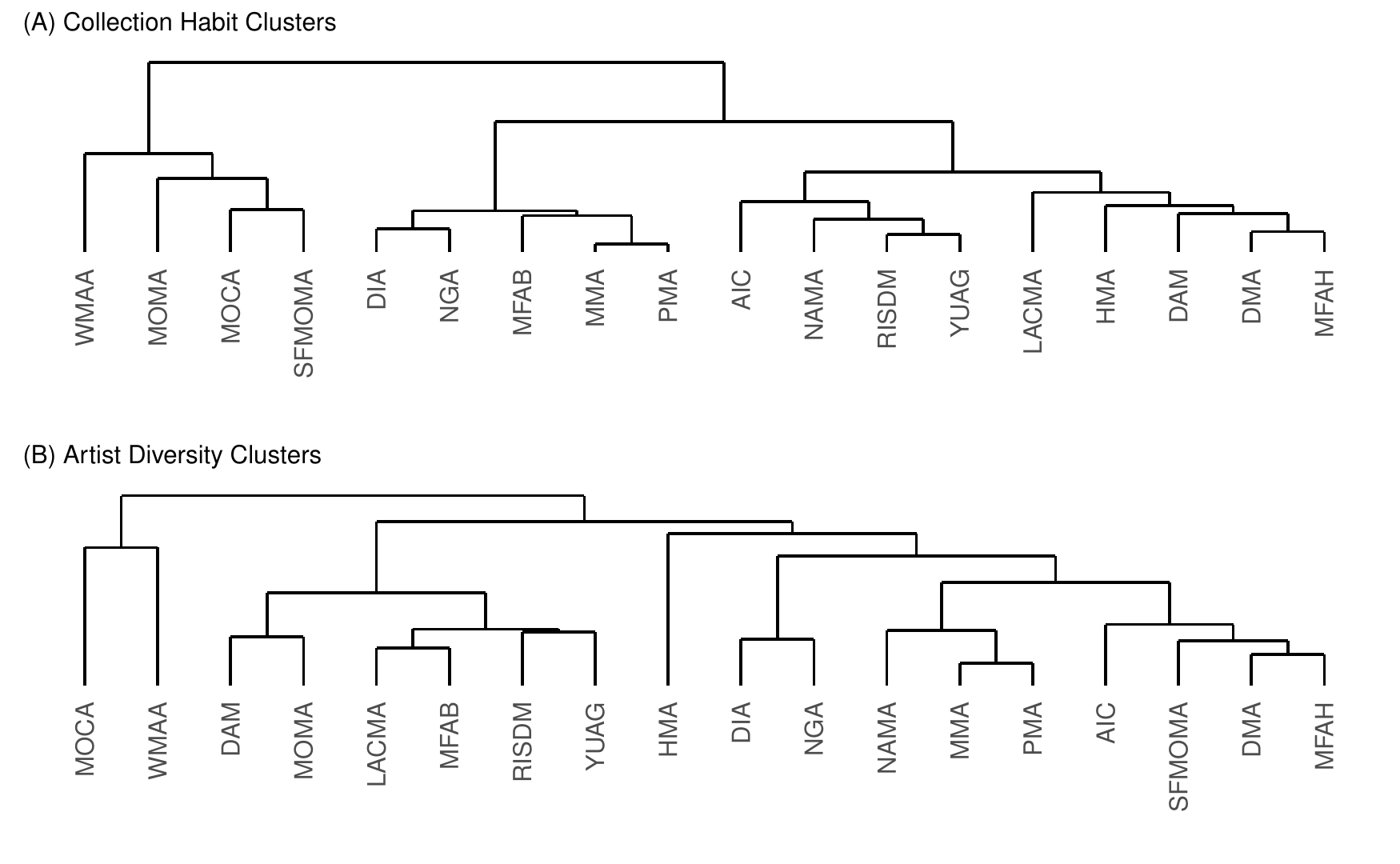}
\caption{Groupings of museums determined by clustering on collection mission demographics and on artist diversity demographics. We use agglomerative hierarchical clustering with average linking under the maximum distance to form clusters. See notes of Table~1 of main text for museum abbreviations. (A) Museum collection mission clusters based on geographic origin and average artist birth year in Table~\ref{Table3}. (B) Museum diversity clusters based on the gender and ethnicity demographics in Table~\ref{Table2}.\label{Fig1}}
\end{adjustwidth}
\end{figure}

We characterized each museum as a point in the six-dimensional space of coordinates, with average birth year values scaled to span the unit interval to be commensurate with proportions for clustering. We ran hierarchical agglomerative clustering with average linking using the maximum distance to obtain clusters. See Fig~\ref{Fig1}(A) for the dendogram. We cut the dendogram at five clusters, which constitute the groupings of Table~\ref{Table1}. Using a similar clustering procedure, we performed a separate grouping of museums based on artist diversity, by characterizing each museum as a point in the five-dimensional space of the gender and ethnicity proportions from Table~\ref{Table2}. We cut the dendogram to obtain three diversity clusters; see Fig~\ref{Fig1}(B).

\begin{table}
\begin{adjustwidth}{-2.25in}{0in} 
\caption{Demographics reflective of museum collection missions, namely geographic origin and artist birth year. We give the observed percentage of artists at each museum confidently inferred to have origin in the following regions: Africa; Asia and the Pacific; Europe; Latin America and the Caribbean; North America; and West Asia. These are the GEO3 regions utilized by the United Nations. We also give the average birth year of artists within the museum (based on their decade-rounded birth years). The table is arranged by collection mission groups determined through cluster analysis; see main text and Fig~\ref{Fig1}. \label{Table3}}
\setlength\tabcolsep{3pt}
\small
\begin{tabular*}{\linewidth}{@{\extracolsep{\fill}}clccccccc}
\hline
Group & Museum & Africa (\%) & Asia (\%) & Europe (\%) & Lat. Am. (\%) & North Am. (\%) & W. Asia (\%) & Avg. Birth Year \\
\hline
{\multirow{5}{*}{1}}&DIA&0.2&2.9&59.5&0.6&36.9&0.0&1802\\
&MMA&0.2&9.5&63.6&0.8&25.7&0.2&1804\\
&MFAB&0.0&16.3&51.2&1.9&30.6&0.0&1803\\
&NGA&0.0&0.9&56.9&0.0&42.2&0.0&1813\\
&PMA&0.4&7.5&61.9&1.9&28.3&0.0&1806\\\hdashline
{\multirow{4}{*}{2}}&AIC&0.0&6.5&56.5&1.4&35.5&0.0&1836\\
&NAMA&0.0&9.7&37.4&0.9&51.8&0.2&1850\\
&RISDM&0.0&13.5&44.2&3.6&38.6&0.2&1849\\
&YUAG&0.0&14.1&39.7&1.9&44.1&0.2&1851\\\hdashline
{\multirow{5}{*}{3}}&DMA&0.2&4.4&45.1&1.4&48.9&0.0&1886\\
&DAM&0.5&8.4&29.7&3.1&58.1&0.3&1886\\
&HMA&2.5&0.3&37.8&0.8&58.6&0.0&1866\\
&LACMA&0.4&17.4&44.4&2.4&35.5&0.0&1885\\
&MFAH&0.4&4.4&38.6&4.0&52.5&0.2&1891\\\hdashline
{\multirow{3}{*}{4}}&MOCA&0.5&5.9&22.3&4.0&67.3&0.0&1949\\
&MOMA&1.0&10.5&47.6&3.1&37.8&0.0&1921\\
&SFMOMA&1.3&7.2&32.8&3.8&55.0&0.0&1929\\\hdashline
{\multirow{1}{*}{5}}&WMAA&0.0&2.1&11.1&1.9&84.7&0.2&1932\\\hdashline
&\textbf{Overall}&\textbf{0.4}&\textbf{8.7}&\textbf{44.0}&\textbf{2.1}&\textbf{44.6}&\textbf{0.1}&\textbf{1863}\\
\hline
\end{tabular*}
\normalsize
\end{adjustwidth}
\end{table}

\begin{figure}
\begin{adjustwidth}{-2.25in}{0in}
\centering
\includegraphics[width=18cm]{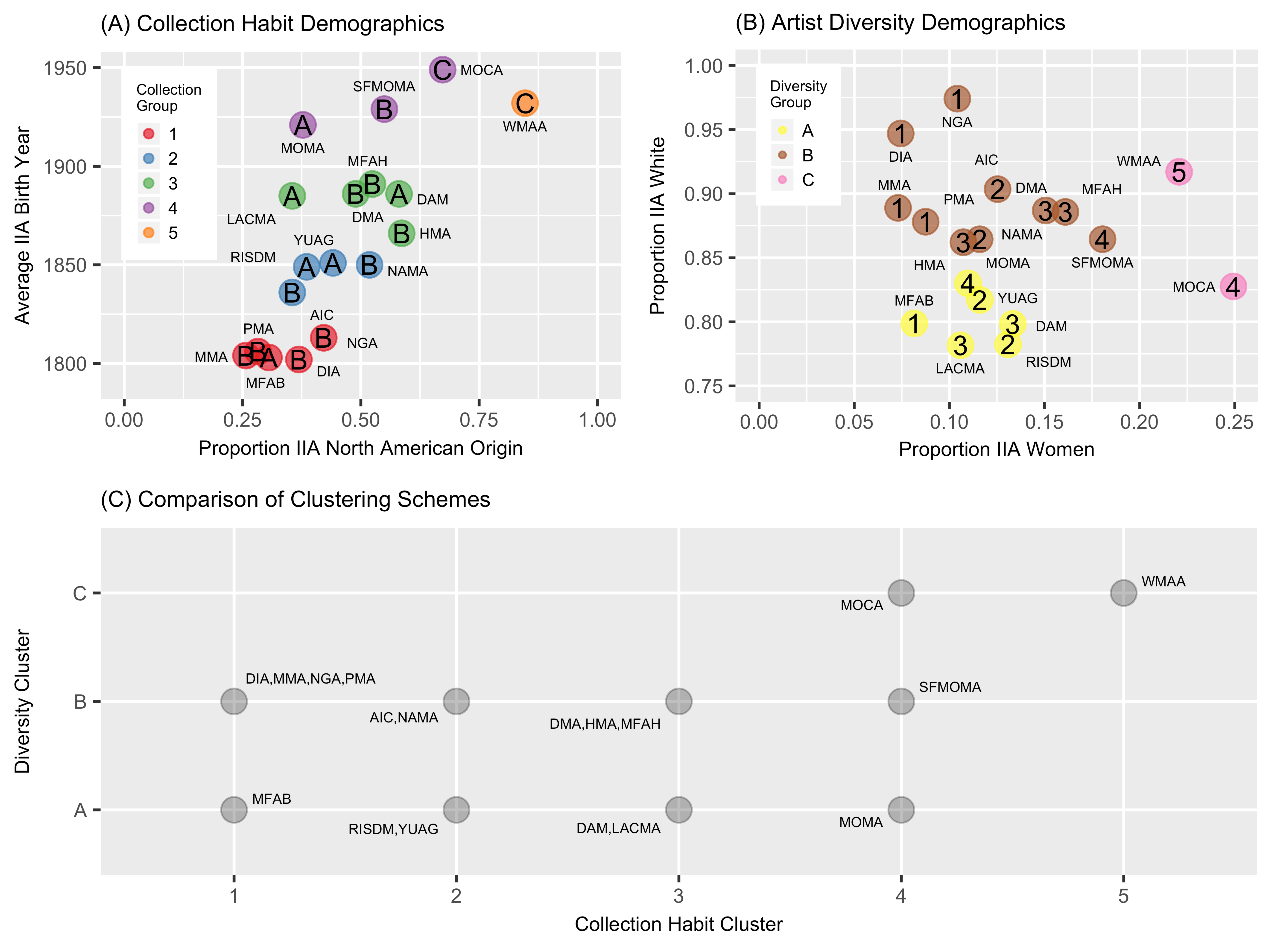}
\caption{Groupings of museums determined by clustering on collection mission demographics and on artist diversity demographics. We use agglomerative hierarchical clustering with average linking under the maximum distance to form clusters. See notes of Table~\ref{Table1} for museum abbreviations.  Museums that have similar collection missions can have markedly different levels of gender and ethnic diversity, and vice versa. (A) Clustering based on demographics indicative of museum collection missions, namely the geographic regions and average birth year as shown in Table~\ref{Table3}. For ease of visualization, we present this six-dimensional data in a simpler format, showing proportion of artists with geographic origin in North America and average birth year of artists. Clusters are color coded and numbered 1 through 4; these are the group numbers in Tables~\ref{Table1}, ~\ref{Table2},~and~\ref{Table3}. The letters inside each circle identify the museum's diversity cluster. (B) Clustering based on demographics indicative of museum artist diversity, namely proportion of women and of each ethnic category, as given in Table~\ref{Table2}. For ease of visualization, we present this five-dimensional data in a simpler format, showing proportion of artists  who are women and proportion of artists who are white. Clusters are color coded and labeled A - D. The numbers inside each circle identify the museum's collection mission cluster from panel (A). (C) Comparison of the two clustering schemes, grouping museums by collection mission cluster on the horizontal axis and diversity cluster on the vertical axis. There is no particular relationship between the two schemes. \label{Fig2}}
\end{adjustwidth}
\end{figure}

Because it is difficult to visualize the museums in the high dimensional spaces mentioned above, we plot the clusters in a simplified setting. Fig~\ref{Fig2}(A) shows the data projected into a plane chosen for easy interpretability; we plot each museum according to its proportion of North American artists and its average artist birth year. Colors indicate the collection mission cluster. The letters inside each circle indicate the museum's diversity cluster. Fig~\ref{Fig2}(B) plots each museum according to its proportion of women artists and white artists. Colors indicate the diversity cluster. The numbers inside each circle indicate the museum's collection mission cluster.

There is not a strong connection between these two different clustering schemes. Clusters with respect to collection mission do not persist when using the different metric of artist diversity. Stated differently, museums with similar collection missions can have quite different diversity profiles, and vice versa. For instance, in  Fig~\ref{Fig2}(A), DAM, DMA, and MFAH have similar collection profiles in terms of geographic origin and time period. However, in Fig~\ref{Fig2}(B), while DMA and MFAH are in the same diversity cluster, DAM is in a different cluster, characterized by substantially fewer white artists, and substantially more men. As another example, in Fig~\ref{Fig2}(A), AIC and RISDM have similar collection profiles, and yet in Fig~\ref{Fig2}(B), they are in different diversity clusters, with RISD substantially less white. Conversely to these examples, in Fig~\ref{Fig2}(B), PMA and and HMA have fairly similar diversity profiles, and yet Fig~\ref{Fig2}(A) shows them in different collection clusters, with PMA having a much older and less North American collection. Fig~\ref{Fig2}(C) provides a summary that compares the two clustering schemes, plotting diversity cluster versus collection mission cluster.

\section*{Discussion and Conclusions}

We studied 10,108 individual, identifiable artist records from 18 museums. By combining crowdsourced data, including confidence ratings, we inferred gender, ethnicity, geographic origin, and birth decade for 89\%, 82\%, 83\%, and 79\% of these individuals, respectively. Restricting our attention to individual, identifiable artists, we found that the artists whose works are held in mainstream U.S. museums are predominantly white men, who comprise 75.7\% of all artists in our pool. We also found that museums with similar foci on time periods and geographic regions can have quite different levels of artist diversity in terms of gender and ethnicity.

Our study comes with limitations. First, all statements about artist demographics are limited to individual, identifiable artists. This restriction affects museums differentially. For example, MFAB boasts 85,000 works of art from Egypt, the Near East, Greece, Italy, and other areas. These generally have no identifiable artist. In contrast, MOCA has an identified artist for nearly all works in the collection. Second, our results carry the assumption that there is no demographic bias in the ability of the crowdworkers to make a confident inference. While we validated our data internally and externally, as reported, we saw some signs that it may be harder to identify women artists from information given on the internet. Future work could investigate this hypothesis. Third, we have focused on artists in a museum's catalog; we could strengthen our inferences about diversity if one accounted for the number of works held by each artist, when and how these works were acquired, and whether or not these works are displayed. Lastly, we have reported on \emph{inferred} demographics of artists. Gender, ethnicity, and even geographic origin are characteristics most appropriately expressed by artists themselves. However, because data is necessary to begin an empirical discussion of diversity, we have carried out inferences using the data available.

Our demographic estimates provide museums with a benchmark for diversity in their collections, and could be used to make decisions impacting collection development. Our clustering results expose a very weak association between collection mission and diversity, thereby opening the possibility that a museum wishing to increase diversity in its collection might do so without changing the geographic and/or temporal emphases of its mission.

The higher the percentage of works by individual, identifiable artists present in a collection, the more complete a demographic picture our study reveals. Our study sheds more light on the demographics of Western artists from the early modern period through the present than it does on those from earlier times and/or those from outside of the West. For most artists in the latter group, artist demographics of the type we study have been lost to history. Future work could creatively develop appropriate and feasible demographic measurements of these older and/or non-Western artists. Combined with the methods we have presented here, a broad demographic picture would then be available to any museum interested in empirical self-reflection as it engaged in collection development with an eye towards diversity.

Finally, our methodology can be used to assess diversity on a large scale in settings other than the art world. As mentioned earlier, diversity in professional and academic spheres has increasingly been a subject of inquiry, including, for example, \cite{NitHebAsh2018,GarMitCue2018}. Important studies like \cite{GarMitCue2018} leverage the availability of raw data -- in that case, Facebook users' self-identified genders. Studies like \cite{NitHebAsh2018} depend on the labor of investigators to manually infer the gender of each person admitted to the study. This is heroic work, though the approach limits the size of a study and largely precludes a validation of the investigators' inferences. Our work efficiently uses crowdsourcing to create a large and robust set of demographic inferences and thus opens up the possibility of accurately assessing diversity on large scales in a variety of settings.

\section*{Acknowledgements}

Laurie Tupper provided valuable input on the design of the pilot study used in our work.

\nolinenumbers


\end{document}